\documentclass[prd,preprintnumbers,showkeys,nofootinbib]{article}
\usepackage{bm, graphicx,amsmath,amssymb,epsf}
\usepackage{subfigure}
\usepackage{epsfig}
\usepackage{color}
\newcommand{\tr}{{\rm tr}} 
\newcommand{\mbf}[1]{\mbox{\boldmath $#1$}}

\newcommand{\bk}{\mbf{k}}

\begin{document}

\title{\Large \bf  The topology of the triple Pomeron vertex in 
${\cal N}=4$ SYM}
\author{\large J.~Bartels, M.~Hentschinski, A.-M.~Mischler 
\bigskip \\
{\it  II. Institute for Theoretical Physics, Hamburg University, Germany} \\
 }
\maketitle

\begin{abstract}
  We investigate, within ${\cal N}=4$ SYM, the high energy behavior in
  the triple Regge limit of a six point correlator of $R$-currents.
  Using the leading logarithmic approximation, we sum all diagrams
  whose color lines fit onto the surface of a sphere.  We find three
  distinct classes of graphs, and one of them contains the triple
  Pomeron vertex known from QCD. We present results which, within the
  AdS/CFT correspondence, can be compared with scattering amplitudes
  in the dual string theory.
\end{abstract}

\section{Introduction}

Within the AdS/CFT correspondence \cite{adscft}, which relates ${\cal N}=4$
Supersymmetric Yang Mills quantum field theory (SYM) in four
dimensions to a string theory in an $AdS_5 \otimes S_5$ space,
correlators of $R$-currents provide a useful tool for investigating
the Regge limit, in particular the correspondence between the BFKL
Pomeron in ${\cal N}=4$ SYM and the graviton on the string side.

In the simplest case, the elastic scattering of two $R$-currents, both
ends of the correspondence have been investigated in leading order.
On the gauge theory side, the supersymmetric impact factors consisting
of the sum of a fermion and of a scalar loop in the adjoint
representation of the gauge group have been computed \cite{Bartels:2008zy}, and it has been
verified that the high energy behavior is described by the BFKL
Pomeron \cite{bfkl}. On the string side, the leading
contribution in the zero slope limit is given by the Witten diagram
with graviton exchange \cite{BKMS1}. Beyond the zero slope limit, the
graviton is believed to reggeize \cite{graviton}.

A next step along this line is the six point function of $R$-currents
in the triple Regge limit. In this kinematic limit one expects to see
the triple Pomeron vertex which represents, besides the BFKL kernel,
another fundamental element of high energy QCD: it describes the
splitting of a BFKL Pomeron into two BFKL Pomerons. On the string
side, one expects the graviton self interaction to play the analogous
role.  For QCD - with fermions in the fundamental representation and
with the electromagnetic current in place of the $R$-current - the six
point function at finite $N_c$ has first been studied in
\cite{Bartels:1994jj}. As the main result, the triple Pomeron vertex
has been identified, which by now has been derived in several other
approaches.  In \cite{BEHM} the case of the six point function of
$R$-currents in ${\cal N}=4$ SYM has been studied for finite $N_c$:
whereas the triple Pomeron vertex remains the same as in the
non-supersymmetric case, a new contribution to the six point correlator
appears which results from the adjoint color representation of the
particles and has no counterpart in QCD.

Another line of interest is the integrability \cite{integrable} of the
BKP equation \cite{bkp}: since in the leading logarithmic
approximation there is no difference between QCD and its
supersymmetric generalization, ${\cal N}=4$ SYM, one expects that the
integrability which has been discovered for the large-$N_c$ limit of
QCD in fact is 'inherited' from ${\cal N}=4$ SYM.  The environment
where the large-$N_c$ limit of BKP states can be studied are higher
order current correlators, e.g. the eight point function in a suitable
multi-Regge limit.  Within the AdS/CFT correspondence, the counterpart
of the BKP states and the role of integrability on the string side has
not been addressed at all.

Recently, an attempt has been made to formulate these high energy
calculations  within a topological approach: in the large-$N_c$ limit, the
color structure of scattering amplitudes can be attributed to
surfaces: spheres, planes, cylinders, pair-of-pants, etc. The
simplest examples include, in QCD with fermions in the fundamental
representation, multi-gluon scattering amplitudes in the plane and the
BFKL Pomeron on the cylinder. In \cite{Bartels:2009zc} also the
six point correlator of electromagnetic currents has been studied in
the large-$N_c$ limit by summing diagrams whose color structure lies
on the surface of a pair-of-pants. The study of these color diagrams
provides a new view on the reggeization of the gluon and on the triple
Pomeron vertex: whereas the reggeization can be understood as a
feature of planar QCD, the triple Pomeron vertex requires a non-planar
structure, reminiscent of the non-planar Mandelstam cross diagram.
     
When turning, from QCD with fundamental fermions, to ${\cal N}=4$ SYM
with fermions and scalars in the adjoint representation, one
encounters changes in the topology of the surfaces and in the
structure of color graphs. The first example is the BFKL Pomeron which
lies on the surface of a sphere with zero boundaries.  Apart from this
the form of the impact factor stays the same.  In the present paper we
address in the large-$N_c$ limit a six point correlator of
$R$-currents in ${\cal N}=4$ SYM in the topological approach.
We sum graphs whose color structure belongs to a specific deformation
of a sphere, corresponding to the pair-of-pants investigated in
\cite{Bartels:2009zc}.

We first review the color structure of Feynman diagrams in the
large-$N_c$ limit and define the classes of graphs which we are going
to sum.  Our final result for the large-$N_c$ limit consists of three
terms which represent three distinct classes of color diagrams on the
surface of the deformed sphere.
     
\section{Topology of graphs in ${\cal N}=4$ SYM}

It may be useful to recapitulate the large-$N_c$ limit of QCD with fermions in the fundamental 
representation, and to recall a few features of the classical 
paper of 't Hooft \cite{'tHooft:1973jz}. Starting from the Fierz identity
\begin{align}
  \label{eq:fierz}
    (g^a)^i_j(g^a)^k_l  =  \delta^i_l \delta^k_j - \frac{1}{N_c} \delta^i_j\delta^k_l,
\end{align}
where $(g^a)^i_j$ denote the $SU(N_c)$ generators in the fundamental
representation (with the normalization $\tr(g^ag^b) = \delta^{ab}$),
and from the identity
\begin{align}
  \label{eq:structure_const}
f^{abc}  
&=
\frac{1}{i\sqrt{2}}\big[\tr(g^ag^bg^c) - \tr(g^cg^bg^a) \big],
\end{align}  
one is lead to draw, in the large-$N_c$ limit, color diagrams with the 
following elements:\\
(i) for each quark in the fundamental representation, a single line
with an arrow, indicating the flow from the upper to the lower index,
\begin{align}
  \label{eq:kronecker_delta}
\delta^i_j = \parbox{2cm}{\includegraphics[width=2cm]{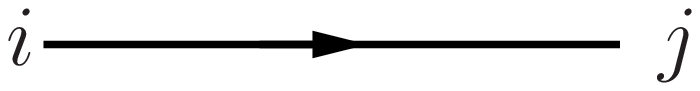}};
\end{align}
(ii) for each inner gluon a double line
\begin{align}
  \label{eq:double_line}
\delta^i_l \delta^k_j= \parbox{2cm}{\includegraphics[width=2cm]{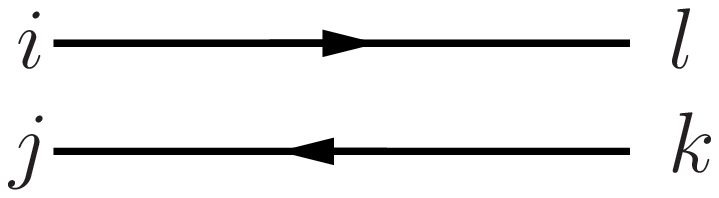}};
\end{align}
(iii) for each triple gluon vertex
\begin{align}
  \label{eq:structure_const2}
f^{abc}  
&=
 \frac{1}{i\sqrt{2}}\big[\tr(g^ag^bg^c) - \tr(g^cg^bg^a) \big]  .
\end{align}  
As a result, each graph turns into a network of double and single
lines. The resulting diagrams represent only the color factors. A
usual Feynman diagram represents both the color factors and the
momentum part. In the double line notation the momentum part has to be
written separately.
   
The double line diagrams can be drawn on a two-dimensional surface
with Euler number $\chi = 2 -2h -b $, where $h$ is the number of
handles of the surface, and $b$ the number of boundaries or holes.  A
closed color-loop always delivers a factor $N_c$, and with the quark
in the fundamental representation of $SU(N_c)$, a closed quark-loop is
$1/N_c$ suppressed, compared to a corresponding gluon-loop, and leads
to a boundary.  For an arbitrary vacuum graph $T$ one arrives at the
following expansion in $N_c$
\begin{align}
  \label{eq:vacuumgraph}
T = \sum_{h,b}^\infty N_c^{2 - 2h -b} T_{h,b}(\lambda)
\end{align}
where
\begin{align}
  \label{eq:thooft}
\lambda &= g^2N_c 
\end{align}
is the 't Hooft-coupling which is held fixed, while $N_c$ is taken to
infinity.

In the expansion Eq.(\ref{eq:vacuumgraph}) which matches the loop
expansion of a closed string theory with the string coupling $1/N_c$,
the leading-$N_c$ diagrams are those that have the
topology of a sphere: zero handles and zero boundaries, $h=b=0$. If
quarks are included, the leading diagrams have the topology of a disk,
i.e. the surface with zero handle and one boundary, $h=0, b=1$, 
fits on the plane, with the boundary as the outermost edge.
Diagrams with two boundaries and zero handles can be drawn on the
surface of a cylinder, those with three boundaries on the surface of a
pair-of-pants. Boundaries are also  obtained by removing, from the
sphere, one or more points. Removing one point, one obtains the disk,
which can be drawn on the plane, and by identifying the removed point
with infinity, the graphs can be drawn on the (infinite) plane.
Removing two points we obtain the cylinder and so on. By definition,
the expansion Eq.(\ref{eq:vacuumgraph}) is defined for vacuum graphs.
However, from the earliest days on \cite{'tHooft:1974hx}, the large-$N_c$ limit has been also applied to the scattering of colored
objects.  In order to consider the topological expansion of an
amplitude with colored external legs, one needs to embed it into a
vacuum graph which then defines the topological expansion of an
amplitude with colored external legs.

\begin{figure}[tbp]
\centering
\includegraphics[height=5cm]{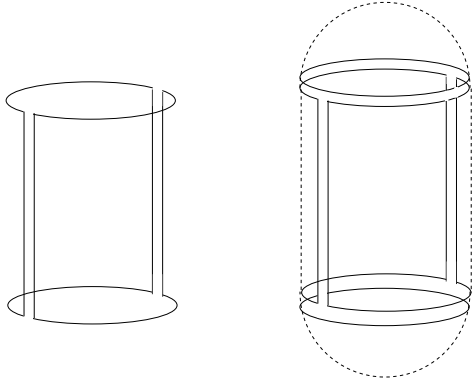}
\caption{\small Cylinder topology for the $ 2 \to 2$ scattering in QCD (left) and in 
${\cal N}=4$ SYM (right)}
\label{cylinder}
\end{figure}

\begin{figure}
\centering
\includegraphics[width=4cm]{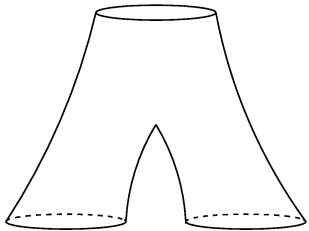}\\
\caption{ \small Pair-of-pants topology  
}
\label{trousers}
\end{figure}

\begin{figure}[ht]
\centering
\includegraphics[height=5cm]{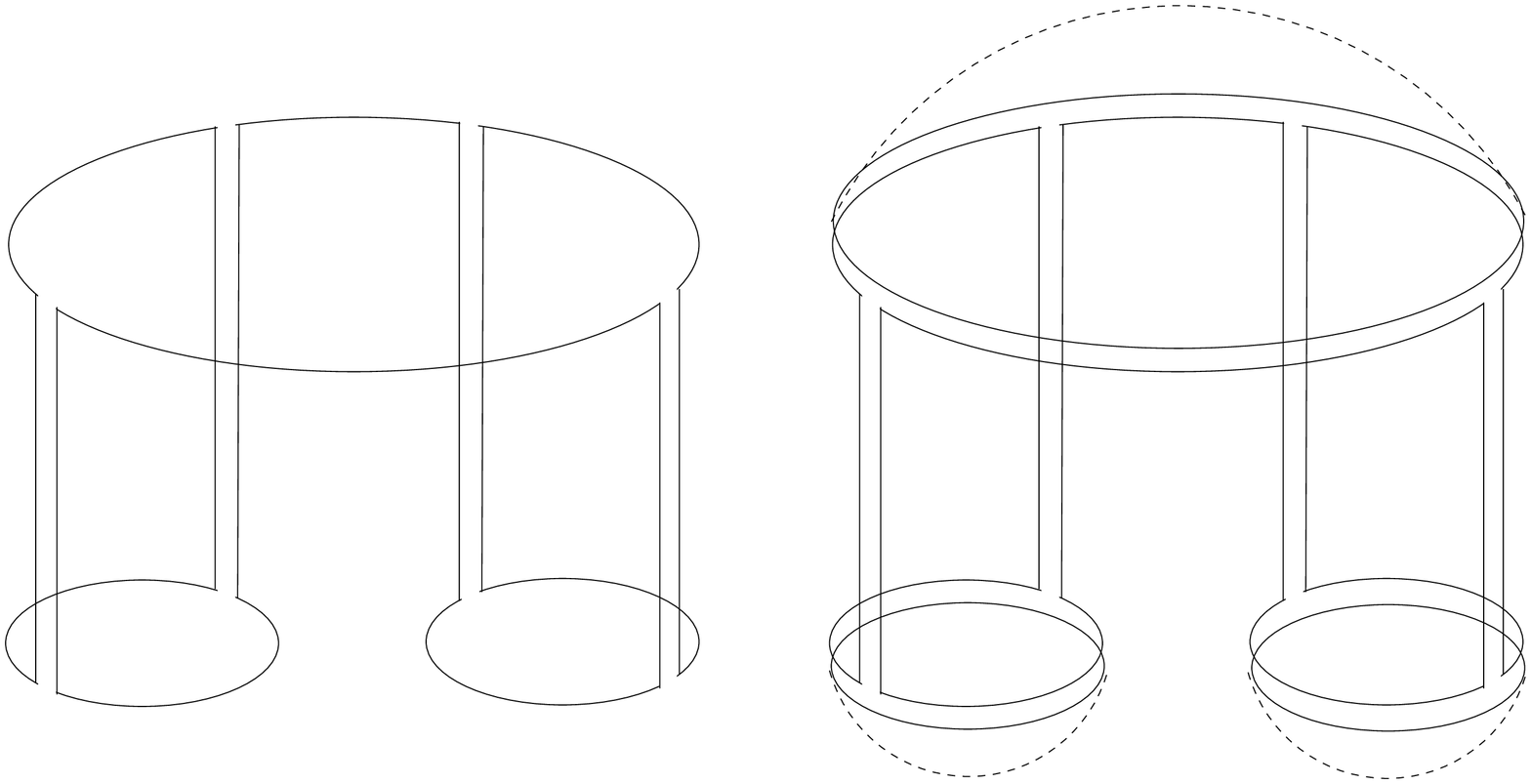}
\caption{ \small Pair-of-pants topology for the $3 \to 3$ amplitude in QCD (left) 
and in ${\cal N}=4$ SYM (right)}
\label{pantsQCDSYM}
\end{figure}

In the large-$N_c$ limit of high energy QCD, quark scattering
amplitudes are drawn on a plane; in particular, one can show that the
BFKL bootstrap condition is satisfied on the plane (zero handles, one
boundary). Next, for the BFKL Pomeron the color diagrams lie on the
surface of a cylinder (zero handles, two boundaries),
(Fig.\ref{cylinder}, left), and the triple Pomeron vertex is obtained
from diagrams which fit on the pair-of-pants surface (zero handles,
three boundaries), (Fig.\ref{trousers} and Fig.\ref{pantsQCDSYM}, left).


Let us now turn to ${\cal N}=4$ SYM where fermions and scalars belong
to the adjoint representation and therefore are represented by double
lines, in the same way as the gauge bosons.  Now fermion loops no
longer define boundaries and therefore cannot be used to define separate
topologies.  As an example, in QCD with fundamental quarks the
diagrams of the BFKL Pomeron fit onto the surface of a cylinder
(Fig.\ref{cylinder}, left): the closed quark loops at the top and at
the bottom define the two boundaries. In ${\cal N}=4$ SYM with adjoint
fermions and scalars, the top and the bottom obtain double lines
(Fig.\ref{cylinder}, right) and turn into caps, as a results of which
the cylinder turns into a (stretched) sphere (zero handles, zero
boundaries).

An analogous result holds for the six point function of external
currents in the triple Regge limit.  For QCD with fundamental quarks
the sphere has three boundaries (for each impact factor, the closed
fermion loop defines a boundary), (Fig.\ref{pantsQCDSYM}, left) and it
has been shown in \cite{Bartels:2009zc} that the color diagrams fit on
the surface of a pair-of-pants (Fig.\ref{trousers}).  When switching
to ${\cal N}=4$ SYM where fermions and scalars belong to the adjoint
representation, the closed lines of the upper and lower impact factors
turn into double lines, and the boundaries of the pair-of-pants are
replaced by caps, (Fig.\ref{pantsQCDSYM}, right).
As a result, we again arrive at a sphere, shaped as a pair-of-pants ,
and we are asked to sum, for the triple Regge limit, diagrams which
lie on the surface of this body. As we shall see in the following,
there are three distinct classes of diagrams: two of them are the same
as in (non-supersymmetric) QCD, whereas the third one
is new and has no analog in QCD.\\

\section{Selection of diagrams}
 
\begin{figure}
\centering
\includegraphics[width=9.5cm]{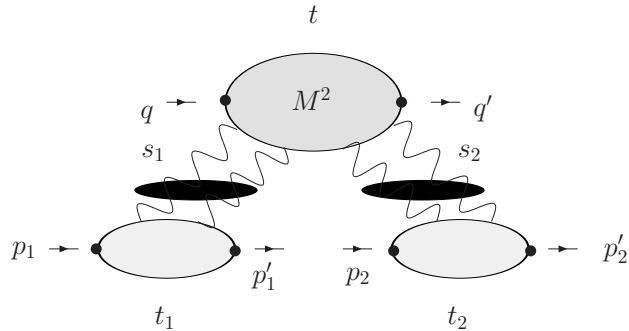}
\caption{ \small Kinematics of a six point amplitude} 
\label{6point}
\end{figure}

Six point amplitudes depend on three energy variables, $s_1=(q+p_1)^2$, $s_2=(q'+p_2')^2$ and $M^2=(q+p_1-p_1')^2$. The momentum transfer variables are $t=(q-q')^2$, $t_1=(p_1-p_1')^2$ and $t_2=(p_2-p_2')^2$. The kinematics of a six point amplitude with external $R$-currents is illustrated in Fig.\ref{6point}.
Then the triple Regge limit is given by
\begin{equation}
\label{tripleRegge} 
s_1=s_2\gg M^2 \gg t,t_1,t_2.
\end{equation}
In the triple Regge limit there is an easy way of computing six point
amplitudes of $R$-currents, namely we take the triple energy
discontinuity in $s_1$, $s_2$ and $M^2$.  In lowest order the main
ingredients of a diagram are three impact factors and four $t$-channel
gluons. The impact factors consist of a fermion and a scalar loop and
represent the coupling of the $t$-channel gluons to the external
$R$-currents. The insertion of the $R$-currents is symbolized by the
small black dots in Fig.\ref{6point}. Higher order corrections are
taken into account by the production of real gluons in the multi Regge
kinematics.  In order to form color singlet states which at the top
couple to the upper $R$-current and at the bottom to two $R$-currents,
one obtains diagrams with four $t$-channel gluons at the lower end and
two, three or four $t$-channel gluons at the upper end. In particular
$t$-channel gluons can emerge from produced real gluons and the number
of $t$-channel gluons can change inside the diffractive system.

We now discuss in the double line notation the relevant diagrams which
contribute to the triple Regge limit of the six point correlators of
$R$-currents. We will closely follow the discussion of
\cite{Bartels:2009zc}, and we will use the same notation. In the
following double line diagrams we do not show the attached
$R$-currents, we only consider the gluons and adjoint particle loops.

We begin with a comment on the cylinder in QCD, (Fig.\ref{cylinder},
left). When replacing, at the top of the cylinder, the fermion in the
fundamental representation by an adjoint fermion, we simply draw,
above the already existing color line of the fermion, an additional
closed color loop which generates an additional factor $N_c$. The
double line notation now also includes scalar loops. The two
$t$-channel gluons are attached to the same color line of the closed
loop, either to the lower line or to the upper one as shown in
Fig.\ref{cylinder}, right. Diagrams where the two $t$-channel lines
are attached to different lines lose this additional factor $N_c$ and
are suppressed. As a result of this simple observation, the
contribution of the adjoint fermions to the impact factors is
proportional to $N_c$ times that of a fundamental one\footnote{This is
  nothing else but the consequence of the different normalizations of
  generators. In the fundamental representation we use $\mbox{tr}(t^a
  t^b)=\delta^{ab}$, whereas in the adjoint representation we have
  $\mbox{tr}(T^a T^b)= N_c \delta^{ab}$. Note that our normalization
  of the fundamental generators deviates by a factor 2 from the
  standard normalization $\mbox{tr} (\tau^a\tau^b)=\delta^{ab}/2$.}.
In ${\cal N}=4$ SYM, the four point correlator is of the form $N_c^2
A_{2 \to 2}(\lambda)$, whereas in non-supersymmetric QCD it is of the
form $N_c^0 A_{2 \to 2}(\lambda)$.

\begin{figure}[bh] \centering
    \begin{minipage}{.9\textwidth}
    \parbox{3cm}{\center   \includegraphics[width=2cm]{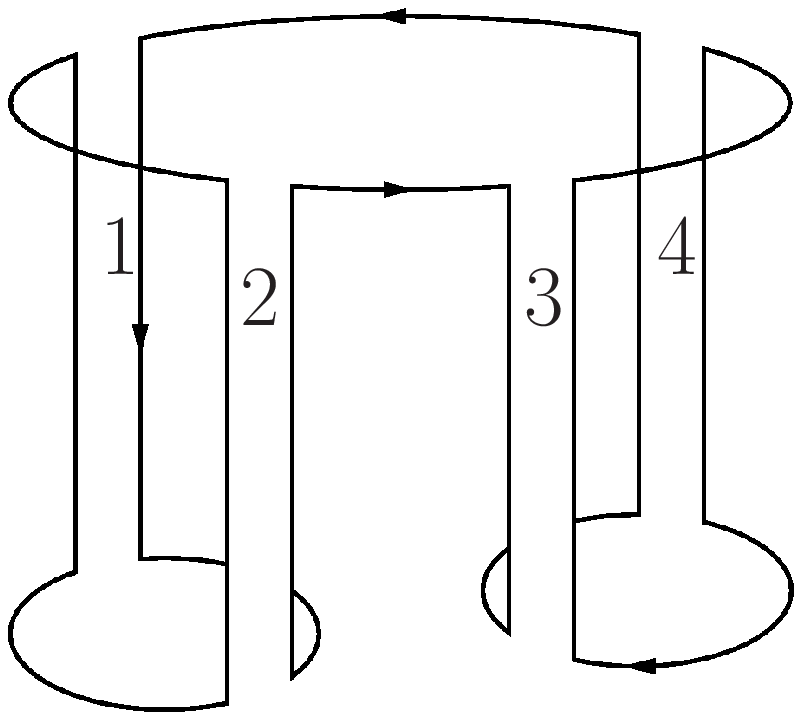}} 
    \parbox{3cm}{\center   \includegraphics[width=2cm]{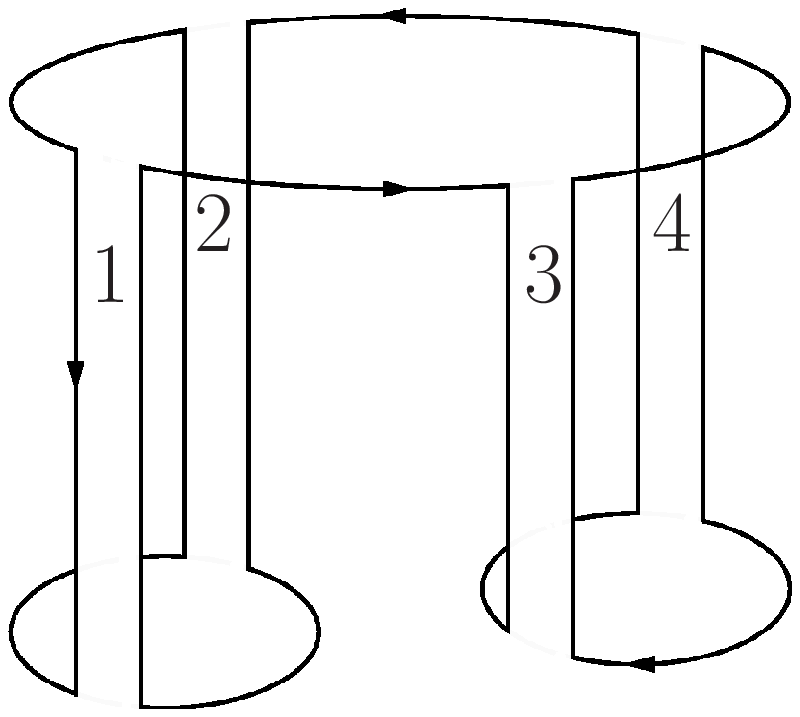}}  
    \parbox{3cm}{\center   \includegraphics[width=2cm]{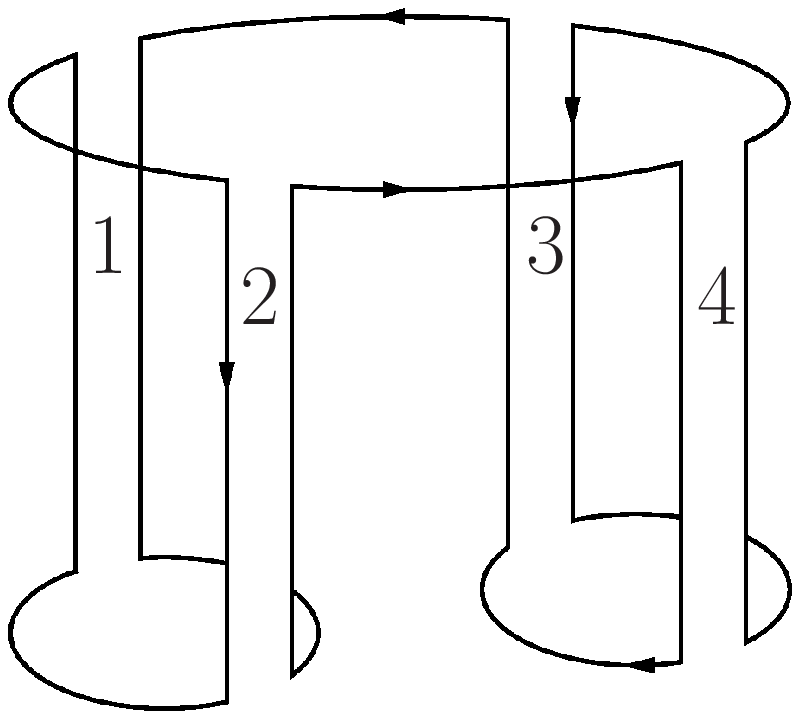}} 
    \parbox{3cm}{\center   \includegraphics[width=2cm]{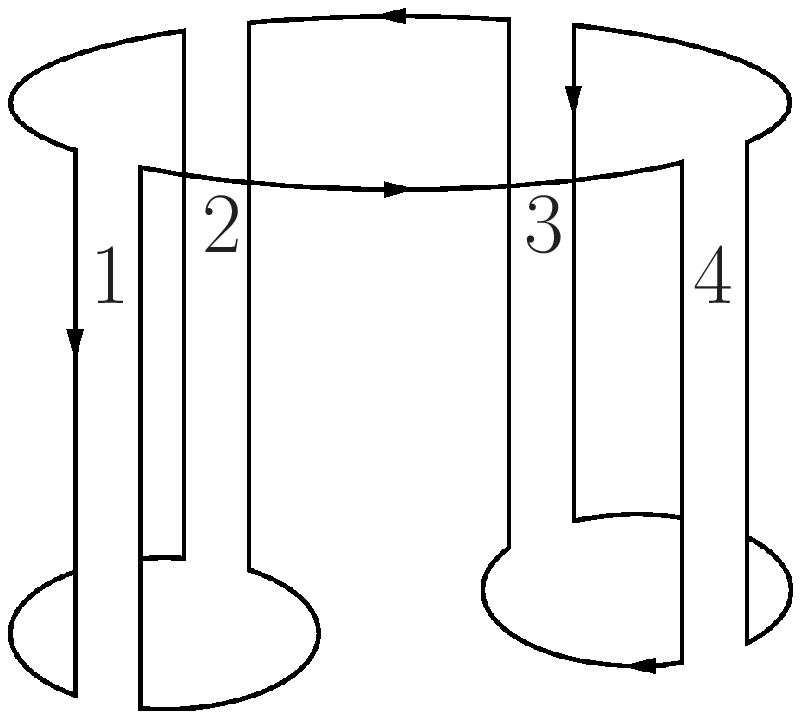}} \\
    \parbox{3cm}{\center $(1234)  $}
    \parbox{3cm}{\center ${(2134)}  $} 
     \parbox{3cm}{\center ${(1243)} $}
    \parbox{3cm}{\center ${(2143)}  $}
  \end{minipage}
  \caption{\small The four different orderings of color factors of the
    Born-term in QCD with fermions in the fundamental representation of $SU(N_c)$}
  \label{fig:born}
\end{figure}
               
Let us now turn to the six point function. In QCD in lowest order the
four $t$-channel gluons couple to the upper fermion or scalar loop in
all possible ways, all together there a 16 different diagrams. A
closer look shows that we have inside the 16 diagrams four different
orderings of color matrices. For the lowest order diagrams in the
non-supersymmetric QCD case, the four different structures are
illustrated in Fig.\ref{fig:born}.

\begin{figure}[tbp]
\centering
\subfigure[]{{\label{Rpantsold}}\includegraphics[height=4cm]{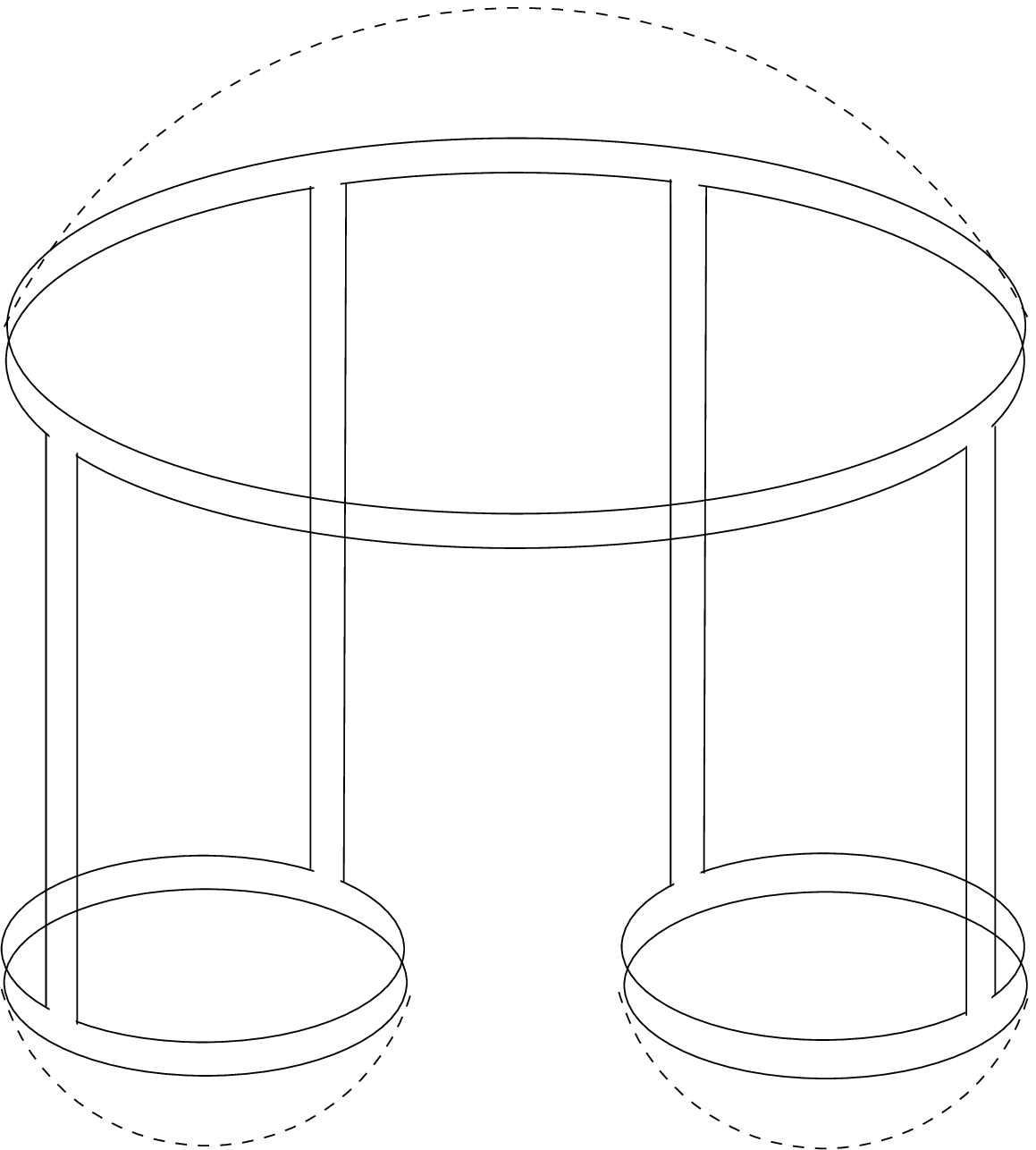}}
\hspace{1cm}
\subfigure[]{{\label{Rpantsnew}}\includegraphics[height=4cm]{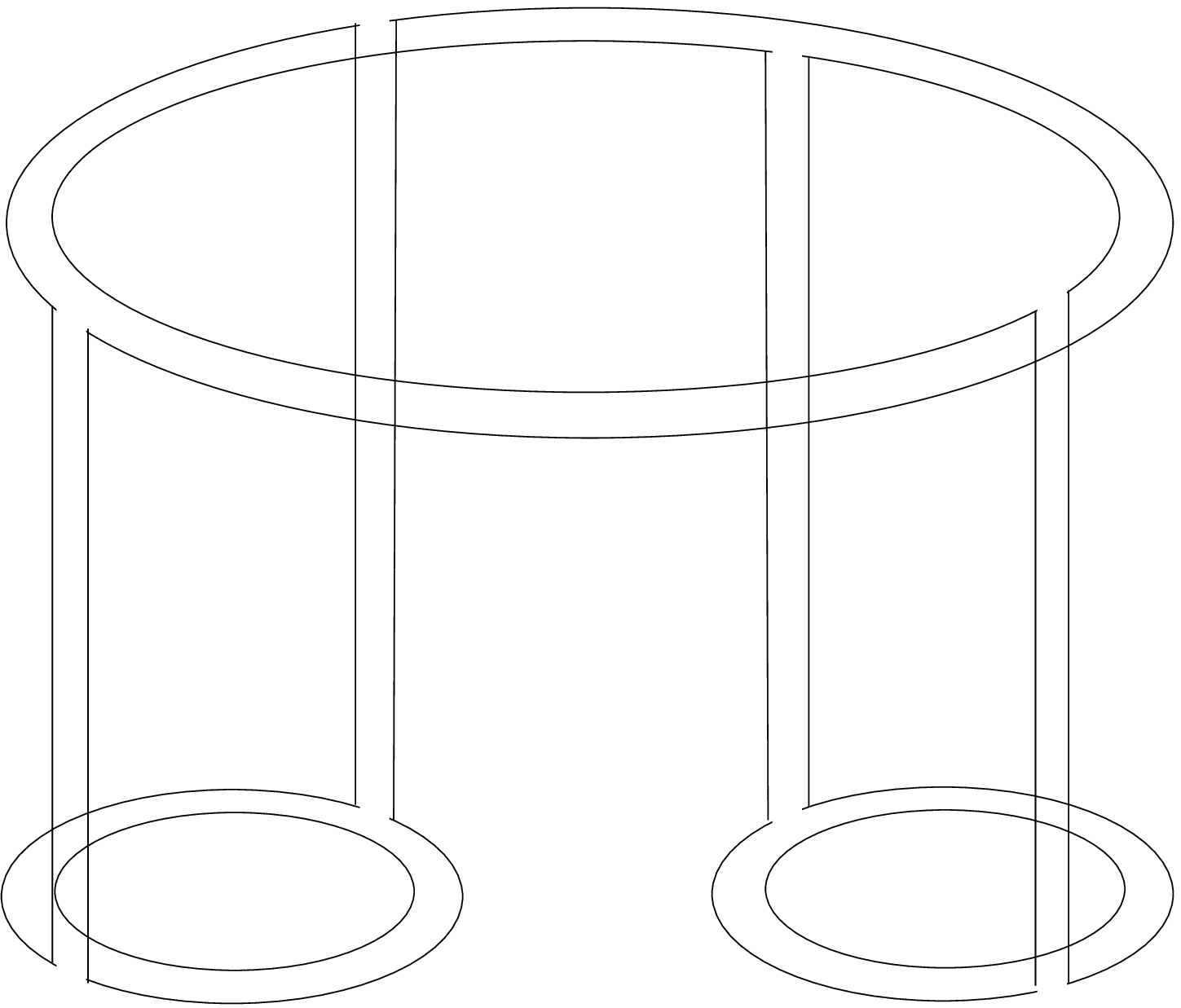}}
\caption{ \small Pair-of-pants topology for the $3 \to 3$ amplitude in ${\cal N}=4$ SYM: (a) a color configuration already present in QCD, and  (b) a new one which exists only for adjoint particles}
\label{Rpants}
\end{figure}

Switching to ${\cal N}=4$ SYM, we simply perform, for each of the
three impact factors, the substitution we have just described for the
BFKL cylinder, and we obtain the additional factor $N_c^3$, leading to
a result of the order $N_c^2 \lambda^4$. Whereas in QCD the analogous
lowest order graphs fit on the surface of a pair-of-pants in
Fig.\ref{trousers}, the diagrams now have the shape of a deformed
sphere as shown once again in Fig.\ref{Rpantsold}. Here both gluons
cylinders are coupled to the same line of the upper loop.

\begin{figure}[bp]
\centering
\includegraphics[height=4cm]{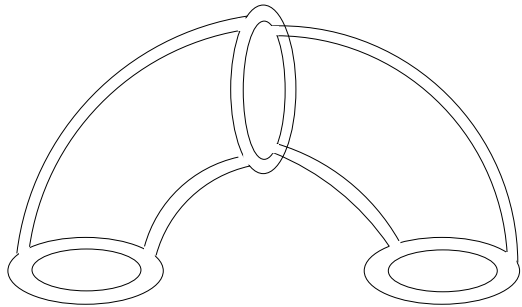}
\caption{ \small An alternative way of drawing Fig.\ref{Rpantsnew}} 
\label{Sausage}
\end{figure}

A closer inspection shows that, in addition to Fig.\ref{Rpantsold},
another configuration is possible: without losing a factor $N_c$, we
can attach one of the cylinders to the outer loop, the other one to
the inner loop (Fig.\ref{Rpantsnew}).  This additional piece in the
four gluon impact factor which has no counterpart in the fundamental
representation, has first been found in \cite{BEHM}.  An alternative
way of drawing this graph is shown in Fig.\ref{Sausage}.

In higher order $\lambda$, several possibilities arise. We briefly
summarize the discussion given in \cite{BEHM}. The general structure
of the diagrams is the following: At the upper impact factor we start
with a $t$-channel state with two, three or four gluons. The
propagation of the $t$-channel gluons is described by the BKP
equations and transition between different states by vertices. We can
have $2\rightarrow 2$, $2\rightarrow 3$, or $2\rightarrow 4$ vertices.
There is always a lowest interaction between the gluons defined by the
$M^2$-discontinuity below which the upper cylinder breaks up into two
disconnected ones. After this branching vertex the gluons interact
only pairwise according to the BFKL equation and are coupled to the
two impact factors of the $R$-currents at the bottom.

\begin{figure}[tb]
\centering
\includegraphics[height=5cm]{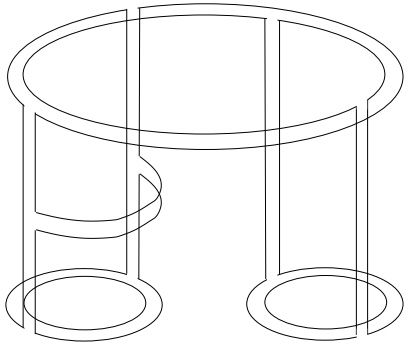}
\caption{ \small Example of a next-to-leading order diagram of Fig.\ref{Rpantsnew}} 
\label{NLO}
\end{figure}

Starting with the one-loop corrections to Fig.\ref{Rpantsnew}, one
realizes that it is only possible to insert $2 \rightarrow 2$
transitions inside the two cylinders: an example is shown in
Fig.\ref{NLO}.
In particular, any rung connecting the two cylinders loses a power of
$N_c$. As a result, this class of diagrams simply consists of two BFKL
Pomerons coupled to the four gluon impact factor, and the resulting
amplitude is of the form $N_c^2 A_{3 \to 3}(\lambda)$. This class of
diagrams will be named 'direct': the two BFKL Pomerons couple directly
to the upper impact factor.  As discussed in \cite{Bartels:2009zc}, on
the cylinder each gluon rung comes in two different ways, one in front
of the cylinder, the other one on the backside. This observation also
applies to the ${\cal N}=4$ SYM case.

\begin{figure}[b]
\centering
\parbox{12cm}{\center \includegraphics[height=4cm]{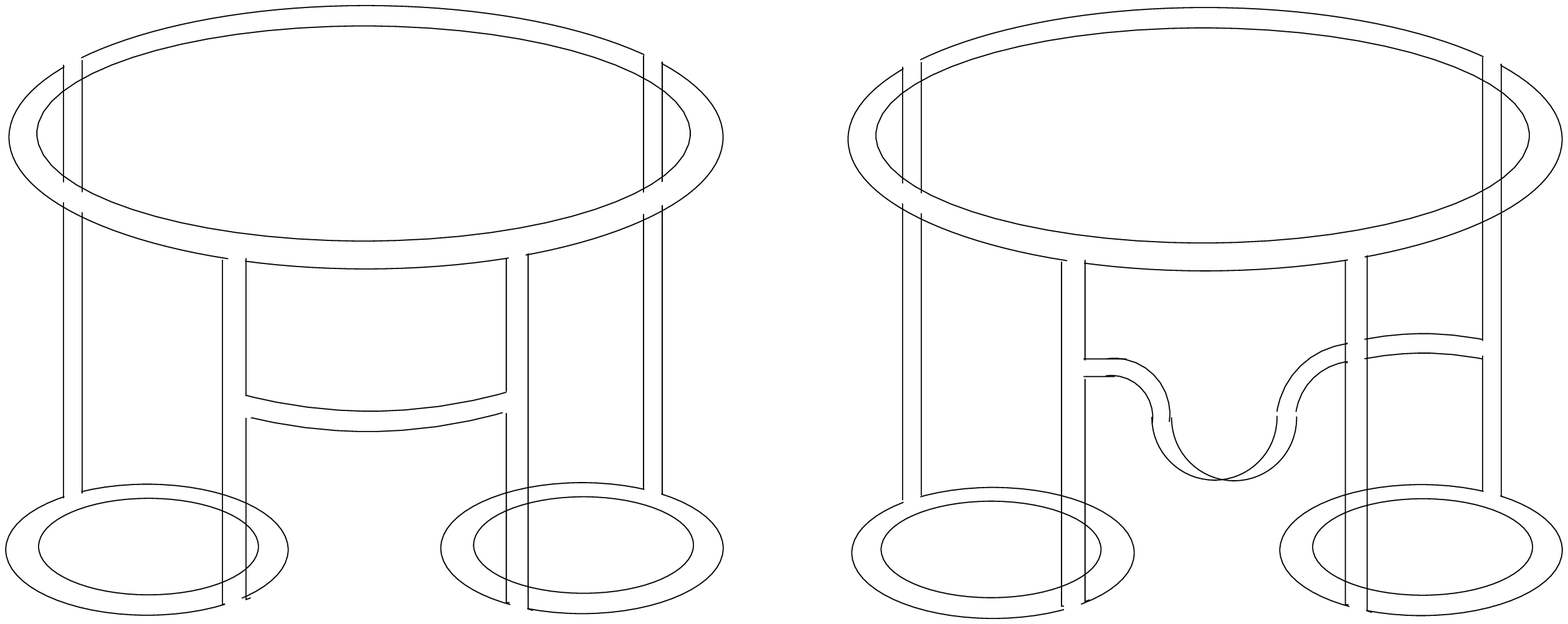}} \\
\parbox{6cm}{\center (a)}\parbox{6cm}{\center (b)}
\caption{ \small Two classes of diagrams: planar graphs (a) and non-planar graphs (b)}   
\label{planarandnonplanar}
\end{figure}
Returning to the diagram in Fig.\ref{Rpantsold}, that is already
present in non-supersymmetric QCD, insertion of one additional rung
opens two distinct classes of graphs: examples are given in
Fig.\ref{planarandnonplanar}, and it is suggestive to name them as
'planar' and 'non-planar', respectively.  By definition, planar graphs
have the property that, by contracting closed color loops, they can be
reduced to the ${\cal N}=4$ SYM version of the graphs in
Fig.\ref{fig:born}. For the non-planar ones, this is not possible.
Beginning with the graph shown in Fig.\ref{planarandnonplanar}b, there
is a new class of diagrams which cannot be deformed into planar
graphs.

Altogether we have to distinguish three different types of diagrams:
the direct, the planar, and the non-planar diagrams.  In the first
case, the direct diagrams, the lowest interaction between the two
cylinders, which defines the value of the diffractive mass $M^2$, is
the upper impact factor itself. The four gluons couple directly the
upper loop without interaction between the two disconnected BFKL
Pomerons.
  
The second type of diagrams are planar diagrams: At the upper loop
they start with two, three, or four $t$-channel gluons.  These gluons
undergo transitions by $2\rightarrow 2$, $2\rightarrow 3$, or
$2\rightarrow 4$ vertices, respectively. One of these transitions is
the branching vertex below which we always have four gluons but each
two gluons only interact pairwise after the branching vertex. They
form once again the two non-interacting BFKL Pomerons.

The last possibility are non-planar diagrams. At the upper loop they
can also start with two, three, or four $t$-channel gluons and the
structure above the branching vertex is the same as for planar
diagrams. But the branching vertex itself now provides a non-planar
structure. Below this non-planar vertex the known disconnected BFKL
cylinders show up.

\section{Analytic expressions}

Let us now turn to analytic expressions. It is convenient to use the
analytic representations of multi-particle amplitudes. A detailed
discussion can be found in \cite{Brower:1974yv}. We restrict ourselves
to those contributions which have a non-vanishing discontinuity in
$M^2$. In the triple Regge limit, Eq.(\ref{tripleRegge}), we have for
the $3\rightarrow 3$ amplitude:
\begin{align}  
T_{3 \to 3}(s_1, s_2, M^2| t_1,t_2,t)=
\frac{s_1s_2}{M^2} \int  \frac{d\omega_1 d\omega_2 d\omega}{(2\pi i)^3}
&
 s_1^{\omega_1}{s_2}^{\omega_2} (M^2)^{\omega-\omega_1-\omega_2}
 \xi({\omega_1})  \xi({\omega_2})  \xi({\omega,\omega_1,\omega_2}) \notag \\
&\cdot
 F(\omega, \omega_1, \omega_2| t, t_1,t_2) +\dots   
\label{eq:tripleregge}
\end{align}
The dots represent three further terms that appear in the triple Regge limit: they do not contribute to the 
$M^2$-discontinuity. The signature factors are given by
\begin{align}
  \label{eq:sig_facs}
            \xi(\omega) &= -\pi\frac{e^{-i\pi\omega}-1}{\sin(\pi\omega)}  
&\mbox{and}& &
 \xi({\omega,\omega_1,\omega_2}) &= -\pi\frac{e^{-i\pi(\omega - \omega_1 -\omega_2)} - 1}{\sin\pi(\omega - \omega_1-\omega_2)}.
\end{align} 
As discussed in the previous section, for the computation we have taken the triple energy discontinuity 
of the amplitude in $s_1$, $s_2$, and  $M^2$:
\begin{equation}
\mathrm{disc}_{s_1} \mathrm{disc}_{s_2} \mathrm{disc}_{M^2} T_{3 \to 3} =
 \pi^3\frac{s_1s_2}{M^2} \int  \frac{d\omega_1 d\omega_2 d\omega}{(2\pi i)^3}
 s_1^{\omega_1}  {s_2}^{\omega_2}  (M^2)^{\omega-\omega_1-\omega_2}
\cdot
 F(\omega,\omega_1, \omega_2 |t_1,t_2,t),
\label{tripledisc}
\end{equation}
which, via a triple Mellin transform, is related to the partial wave
$F(\omega,\omega_1, \omega_2 |t_1,t_2,t)$. For the calculation of the
triple discontinuity we have used unitarity integrals, and the
summation of the diagrams has been performed by means of integral
equations.  These integral equations can be solved and the solution is,
due to the resulting boostrap,  far more simpler than suggested by the
structure of the underlying diagrams.  Details can be found in
\cite{Bartels:2009zc}, and we only describe the results.
\begin{figure}
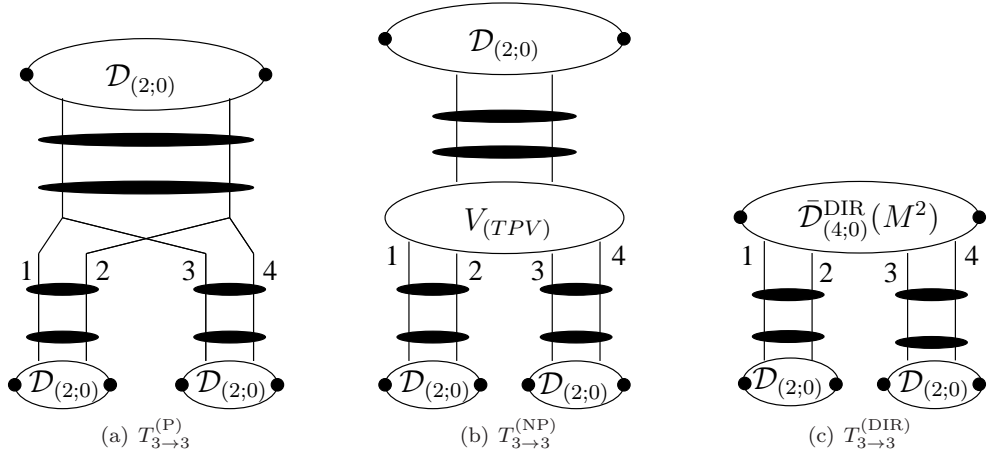

\centering 
\subfigure[   $T_{3 \to 3}^{(\text{P})}$]{{\label{Fplanar}}\input{Fplanar.pstex_t}}
\hspace{1cm}
\subfigure[ $T_{3 \to 3}^{(\text{NP})}$]{{\label{Fnonplanar}}\input{Fnonplanar.pstex_t}}
\hspace{1cm}
\subfigure[$T_{3 \to 3}^{(\text{DIR})}$]{{\label{Fdirect}}\input{Fdirect.pstex_t}}
\caption{ \small The three different parts of the six point function $T_{3\to3}$
}
\label{partialwaves}
\end{figure}

Our result for the sum of the diagrams which fit on the surface of the
deformed sphere (Fig.\ref{Rpants}) is given by the sum of three terms
which sum different classes of diagrams: planar diagrams, non-planar
diagrams, and direct diagrams:
\begin{equation}
T_{3 \to 3}= T_{3 \to 3}^{(\text{P})}+ T_{3 \to 3}^{(\text{NP})} + T_{3 \to 3}^{(\text{DIR})}.
\end{equation}
The three different parts are illustrated in Fig.\ref{partialwaves}.
In the triple Regge limit, Eq.(\ref{tripleRegge}), the partial waves
factorize and consist of several building blocks. In our case we
encounter two-gluon impact factors, BFKL Green's functions, and a
triple vertex which connects them.  The two-gluon impact factors,
$\mathcal{D}_{(2;0)}$, describe the coupling to the external
$R$-currents. For ${\cal N}=4$ SYM they have been computed in
\cite{Bartels:2008zy}. They contain both fermions and scalar in the
adjoint representation\footnote{For the large $N_c$ analysis it is
  convenient to absorb an additional factor $N_c$ into the impact
  factor. With the two-gluon $R$-current impact factor of
  \cite{Bartels:2008zy} given by $\Phi^{aa'} = \delta^{aa'}\Phi$, we
  use in the following $\mathcal{D}_{(2;0)} = N_c\Phi $, which is then
  proportional to $\lambda N_c$.  }.

Let us go into some detail. For the sum of the first two terms we use
the representation Eq.(\ref{eq:tripleregge}) and write
\begin{equation}
F(\omega, \omega_1,\omega_2)=F^{(\text{P})}(\omega, \omega_1,\omega_2)+F^{(\text{NP})}(\omega, \omega_1,\omega_2).
\end{equation}
As seen in Fig.\ref{partialwaves}, the impact factors appear at the
three different ends of the diagrams, and they are connected by BFKL
Green's functions and a triple vertex in the center.  The two terms
differ from each other by the form of the triple vertex: in the second
term, the vertex is due to the non-planar diagrams and coincides with
the triple Pomeron vertex found in QCD. The first term which results
from the planar diagrams is a direct consequence of the reggeization
of the gauge boson. The partial wave has the form:
\begin{align}
  \label{eq:FP}
F^{(\text{P})}(\omega, \omega_1, \omega_2) 
&= 4 
 \mathcal{D}^{(12)}_2(\omega_1) \otimes_{12}
 \mathcal{D}^{(34)}_2(\omega_2)\otimes_{34}
\left(\bar{D}_{(4;0)}(\omega) + 
   \big[\omega - \omega_1 -\omega_2 \big] \frac{\bar{\lambda}}{N_c} V^R \otimes
 \mathcal{D}_{2}(\omega)  \right)
\end{align}
with  $ \bar\lambda={\lambda}/{2} $.
The convolution symbol is defined as
\begin{equation}
\otimes_{12}=\int\frac{d^2{\bf k}_1}{(2\pi)^3{\bf k}_1^2{\bf k}_2^2},
\end{equation}
where ${\bf k}_1$ and ${\bf k}_2$ are the transverse momenta of the
gluons 1 and 2.  In Eq.(\ref{eq:FP}) we have introduced the three
functions $\mathcal{D}_2(\omega)$, $\mathcal{D}_2(\omega_1)$,
$\mathcal{D}_2(\omega_2)$ which combine the three impact factors
$D_{(2;0)}$ with their adjacent BFKL Green's functions.  The subscript
$12$ at the convolution symbol indicates that the two gluon amplitude
$\mathcal{D}^{(12)}_2(\omega_1)$ has to be contracted with the
$t$-channel gluons 1 and 2. Similarly, $\otimes_{34}$ belongs to the
gluons $3$ and $4$.  An analytic expression for the triple vertex $V^R$
can be found in \cite{Bartels:2009zc}.

The second part of the partial wave takes the form
\begin{align}
  \label{eq:F_nonplanar}
F^{(\text{NP})}(\omega, \omega_1, \omega_2) 
&=
4  \mathcal{D}^{(12)}_2(\omega_1) \otimes_{12} \mathcal{D}^{(34)}_2(\omega_2)\otimes_{34}
   \frac{\bar{\lambda}^2}{N_c} V_{(\text{TPV})}  \otimes \mathcal{D}_2(\omega).
\end{align}
The main ingredient here is the triple Pomeron vertex
$V_{(\text{TPV})}$, which as described in \cite{Bartels:2009zc},
Eq.(87), coincides with the large-$N_c$ limit of the QCD result of
\cite{Bartels:1994jj}.  Interesting enough, the first term,
$F^{(\text{P})}$, is present only in the triple Regge limit with fixed
$M^2$. As explained in \cite{BEHM, Bartels:2008ru}, after integration
over $M^2$ and $t_1$ and $t_2$, our six point function can be viewed
as a part of the scattering of the upper $R$-current on a loosely
bound state of the two lower $R$-currents: in this case this part of
the triple vertex disappears and turns into a special contribution to
the initial conditions of the evolution of a BFKL Green's function,
and only the triple Pomeron vertex remains (a detailed discussion of
this mechanism is given in \cite{Bartels:1994jj}, after Eq.(4.14)).

Finally, we have the third part in Fig.\ref{partialwaves}, $T_{3 \to
  3}^{\text{(DIR)}}$, where the two BFKL Green's functions couple
directly to the upper 'unintegrated' impact factor,
$\bar{\mathcal{D}}_{(4;0)}^{\text{DIR}}(M^2)$. Here the dependence
upon $M^2$ is contained inside
$\bar{\mathcal{D}}_{(4;0)}^{\text{DIR}}$, and instead of
Eq.(\ref{eq:tripleregge}) we use
\begin{align}  
  \label{eq:direct}
    T_{3 \to 3}^{\text{(DIR)}}(s_1, s_2, M^2| t_1,t_2,t)=
    s_1s_2 \int  \frac{d\omega_1 d\omega_2}{(2\pi i)^2}
    &
    \left(\frac{s_1}{M^2}\right)^{\omega_1}\left(\frac{s_2}{M^2} \right)^{\omega_2} 
    \xi({\omega_1})  \xi({\omega_2}) \;
    F^{(\text{DIR})}(M^2, \omega_1, \omega_2| t, t_1,t_2).  
\end{align}
The partial wave is given by
\begin{align}
   \label{Fdirfinal}
       F^{(\text{DIR})}(M^2,\omega_1,\omega_2 | t, t_1, t_2)&=
       4\, \mathcal{D}^{(12)}_2 (\omega_1) \otimes_{12} 
       \mathcal{D}^{(34)}_2 (\omega_2)  \otimes_{34}
       \bar{\mathcal{D}}_{(4;0)}^{\text{DIR}}(M^2).
\end{align}
The direct coupling of the amplitudes $\mathcal{D}^{(12)}_2(\omega_1)$
and $\mathcal{D}^{(34)}_2(\omega_2)$ to the upper loop is described by
the 'unintegrated' impact factor
$\bar{\mathcal{D}}_{(4,0)}^{{\text{DIR}}}(M^2)$ which has both right
hand and left hand cuts in $M^2$.  In contrast to the 'normal' impact
factors which are integrated over the mass $M^2$, in this case the
coupling to the upper $R$-current is with fixed $M^2$.  Restricting
ourselves to the forward case with zero momentum transfers
$t=t_1=t_2=0$, i.e. ${\bm k}_1 = -{\bm k}_2 = {\bm {k}}$ and ${\bm
  k}_3 = -{\bm k}_4 = {\bm {k}}'$, we obtain after integration over
the angle $\varphi$ and $\varphi'$ of the vector ${\bm k}$ and ${\bm
  k}'$ respectively,
\begin{align}
   \label{Fdirfinalforward}
       F^{(\text{DIR})}&(M^2,\omega_1,\omega_2 | 0,0,0)=
\notag \\
&=
       4\,
\int \frac{d {\bm k}^2}{(2\pi)^3{\bm k}^4}
\int \frac{d {{\bm k}'}^2}{(2\pi)^3{{\bm k}'}^4}
 \mathcal{D}^{(12)}_2 (\omega_1,{\bm k}^2 ) 
       \mathcal{D}^{(34)}_2 (\omega_2, {{\bm k}'}^2)
       \bar{\mathcal{D}}_{(4;0)}^{\text{DIR}}(\bk^2,{\bk'}^2,M^2) ,
\end{align}
where 
(see also \cite{BEHM}):
\begin{eqnarray}
       \label{eq:unintIFhh}
       \bar{\mathcal{D}}_{(4;0)}^{{\text{DIR}};hh'}(\bk^2,{\bk'}^2,M^2)
       =
       \frac{g^4}{32 }\frac{1}{M^2} \delta_{hh'} 
       \int_0^1 d \alpha\,
       I_v(\alpha, \bk^2,M^2) I_v(\alpha, {\bk'}^2,M^2)
\end{eqnarray}
and 
\begin{eqnarray}
         \label{eq:unintIFLL}
         \bar{\mathcal{D}}_{(4;0)}^{{\text{DIR}};{LL}}(\bk^2,{\bk'}^2,M^2)
         =
         \frac{g^4}{32}Q^2 
         \int_0^1 d \alpha\,\alpha^2 (1-\alpha)^2 
         I_s(\alpha, \bk^2,M^2) I_s(\alpha, {\bk'}^2,M^2).
\end{eqnarray}
Here the dependence on $M^2$ is explicit. $L$ and $h$ denote the
different polarizations of the incoming and outgoing $R$-currents.
The functions $I_v$ and $I_s$ are given by
\begin{eqnarray}
 I_v(\alpha,\bk^2,M^2)
= 
 \left( \frac{Q^2 - M^2}{Q^2 + M^2} - 
\frac{\bk^2 +\alpha (1-\alpha) (Q^2 - M^2)}
{\sqrt{(\bk^2+\alpha(1-\alpha) (Q^2 - M^2))^2 + 4 \alpha^2(1-\alpha)^2 M^2 Q^2}}\right),
\end{eqnarray}   
and 
\begin{eqnarray}
I_s(\alpha,\bk^2,M^2)
= 2 \left( \frac{1}{\sqrt{(\bk^2+\alpha(1-\alpha) (Q^2 - M^2))^2 + 4 \alpha^2(1-\alpha)^2 M^2 Q^2}} 
- \frac{1}{\alpha(1-\alpha)(Q^2 + M^2)}\right).
\end{eqnarray}
A final comment on $F^{(\text{DIR})}$: Compared to $F^{(\text{P})}$
and $F^{(\text{NP})}$ this term is sub-leading for large values of
$M^2$. It therefore  contributes in the region of low-diffractive
mass $M^2$, whereas the region of high-diffractive mass $M^2$ is
governed by $F^{(\text{P})}$ and $F^{(\text{NP})}$.

The expressions listed in this section can directly be compared with
results obtained from the analysis of Witten diagrams in the strong
coupling regime \cite{BKMS2}.

\section{Conclusions}
In this letter we have presented results for the six point $R$-current
correlator in ${\cal N}=4$ SYM in the triple Regge limit. In view of
the AdS/CFT correspondence, which relates the large-$N_c$ limit
of ${\cal N}=4$ SYM to a string theory in $AdS_5$, we have concentrated
on those diagrams which fit on the surface of a sphere. This
generalizes an earlier calculation in non-supersymmetric QCD, where the
leading diagrams fit on the surface of a pair-of-pants.

Our result consists of three terms which correspond to different
classes of color diagrams. Each term is composed of several buildings
blocks: impact factors, triple vertex, and BFKL Green's functions.
They should be compared with corresponding Witten diagrams in the
strong coupling region. Further work along these lines is in progress
\cite{BKMS2}.

\end{document}